\documentstyle[12pt]{article}
\newcounter{abc}

\newcommand{\gE}{\mbox{$\tau$}}
\newcommand{\gM}{\mbox{$h$}}
\newcommand{\oE}{\mbox{${\cal E}$}}

\newcommand{\oM}{\mbox{${\cal M }$}}

\newcommand{\aE}{\mbox{$a_{{\cal E}}$}}
\newcommand{\aM}{\mbox{$a_{{\cal M}}$}}

\newcommand{\LM}{\mbox{$\Lambda _{{\cal M}}$}}
\newcommand{\LMp}{\mbox{$\Lambda _{\cal M}^+$}}

\newcommand{\LE}{\mbox{$\Lambda _{{\cal E}}$}}
\newcommand{\LEp}{\mbox{$\Lambda _{{\cal E}}^+$}}

\newcommand{\ptMstar}{\mbox{$\tilde{p}^\star(x)$}}
\newcommand{\ptMEstar}{\mbox{$\tilde{p}^\star(x,y)$}}
\newcommand{\ptEstar}{\mbox{$\tilde{p}^\star(y)$}}

\newcommand{\pLM}{\mbox{$p_L({\cal M})$}}
\newcommand{\pLME}{\mbox{$p_L({\cal M},{\cal E})$}}
\newcommand{\pLE}{\mbox{$p_L({\cal E})$}}

\newcommand{\mix}{\mbox{$s$}}
\newcommand{\mixp}{\mbox{$r$}}
\newcommand{\dfac}{\mbox{$\frac{1}{1-\mix \mixp}$}}

\begin{document}

\title{Liquid-vapour asymmetry in pure fluids:\\ a Monte Carlo
simulation study}

\author{N. B. Wilding and M. M\"{u}ller\\ {\small Institut f\"{u}r
Physik, Johannes Gutenberg Universit\"{a}t,}\\ {\small Staudinger Weg
7, D-55099 Mainz, Germany}}

\date{July 1994}
\setcounter{page}{0}
\maketitle

\begin{abstract}

Monte Carlo simulations within the grand canonical ensemble are used
to obtain the joint distribution of density and energy fluctuations
$p_L(\rho,u)$ for two model fluids: a decorated lattice gas and a
polymer system. In the near critical region the form of $p_L(\rho,u)$
is analysed using a mixed field finite-size-scaling theory that takes
account of liquid-vapour asymmetry. Field mixing transformations are
performed that map $p_L(\rho,u)$ onto the joint distribution of
critical scaling operators \ptMEstar\ appropriate to the Ising fixed
point. Carrying out this procedure permits a very accurate
determination of the critical point parameters.  By forming various
projections of \ptMEstar , the full universal finite-size spectrum of the
critical density and energy distributions of fluids is also obtained.
In the sub-critical coexistence region, an examination is made of the
influence of field mixing on the asymmetry of the density
distribution.

\end{abstract}

\thispagestyle{empty}
\begin{center}
PACS numbers 64.70F, 05.70.Jk
\end{center}
\newpage

\section{Introduction}

Over the years, finite-size-scaling (FSS) techniques have proved
themselves an indispensable tool for computer simulation
investigations of critical phenomena in model systems, facilitating
accurate estimates of infinite volume quantities from simulations of
finite size. Of the many previous FSS simulation studies that have
been performed \cite{PRIVMAN}, most have focussed on critical
phenomena in lattice-based magnetic spin systems such as the Ising
\cite{BINDER3,FERRENBERG1}, $\phi^4$ \cite{NICOLAIDES}, XY
\cite{GOTTLOB} and Heisenberg models \cite{PECZAK,CHEN}. Among the
specific approaches employed to study such systems, use of the order
parameter distribution function has proved itself one of the most
powerful. The FSS properties of the order parameter distribution are
now routinely employed in studies of magnetic systems, facilitating
both the accurate location of the critical point and the detailed
elucidation of its character \cite{BINDER1}.

Only comparatively recently has attention turned to the task of
applying FSS techniques to the simulation of critical fluids. Work to
date has concentrated on attempting to carry over to fluids the order
parameter block distribution techniques developed in the magnetic
context \cite{ROVERE,BRUCE2,WILDING1}. In the process, however, it has
been found necessary to generalise the FSS equations to take account
of the absence in fluids of the energetic (`particle-hole' \cite{FN1})
symmetry that prevails in magnetic systems such as the Ising model.
Although this reduced symmetry is believed to have no bearing on the
universal critical behaviour of fluids, (which for system with
short-ranged interactions correspond to the Ising universality class),
it has long been recognised that it leads to certain non-universal
effects.  The most celebrated of these is a weak energy-like
singularity in the coexistence diameter on the approach to
criticality, the existence of which constitutes a failure for the law
of rectilinear diameter \cite{SENGERS}.

Within the renormalisation group framework, the critical behaviour of
a given system is characterised by the values of its relevant scaling
fields specifying the location of the effective hamiltonian with
respect to the fixed point \cite{WEGNER}. In models of the Ising
symmetry, these scaling fields are simply identifiable with the
thermodynamic fields, namely the (reduced) temperature and the applied
field. By contrast for fluids, the absence of particle-hole symmetry
implies that the scaling fields comprise {\em mixtures} (i.e. linear
combinations) of the temperature and chemical potential. As a
consequence of this `field mixing', the fluid scaling operators (the
quantities conjugate to the two relevant scaling fields) are also
predicted to differ from those of the symmetric Ising systems.  In the
Ising model, the scaling operators are simply the order parameter
(i.e. magnetisation) and the energy density. In fluids however, they
are expected to be linear combinations of the order parameter
(particle density) and the energy density.

The modified forms of the scaling variables have recently been
incorporated within a FSS theory describing the interplay of energy
and density fluctuations in near-critical fluids
\cite{BRUCE2,WILDING1}. This theory provides a potentially powerful
framework for the detailed simulation study of critical phenomena in
fluids. Hitherto, however, only a limited appraisal of the theory has
been performed. Attention has focused on one predicted manifestation
of field mixing, namely a finite size correction to the limiting form
of the critical order parameter distribution. The existence of this
correction has indeed been confirmed in detailed Monte-Carlo
simulation studies of both the 2D Lennard-Jones fluid \cite{WILDING1}
and a 2D decorated lattice gas model \cite{WILDING2}.  To date,
however, the full extent of the claims embodied in the mixed field FSS
theory have not been closely scrutinised.

In the present paper we address this matter with simulation studies of
two critical fluid systems: a decorated lattice gas model and a
polymer model. The paper is broadly organised as follows. We begin by
providing a short resum\'{e} of the mixed field FSS theory for the
near-critical density and energy fluctuations of fluids. We then
present simulation measurements of the joint distribution of the
density and energy fluctuations $p_L(\rho,u)$ at the liquid-vapour
critical point of a 3D decorated lattice gas model. Field mixing
transformations are performed that map $p_L(\rho,u)$ onto the fixed
point distribution of scaling operators \ptMEstar\ appropriate to the
Ising universality class. Effecting this data collapse yields
estimates of the field mixing parameters that control the degree of
field mixing, the values of which are found to be in excellent
agreement with analytic calculations. Application of field mixing
transformation to \ptMEstar\ are also used to generate the full
universal finite-size spectrum of the density and energy density
distributions of fluids. This analysis reveals, in particular, that
(compared to models of the Ising symmetry) the presence of field
mixing radically alters the limiting (large L) form of the critical
energy distribution.

Consideration is then given to the role of field mixing in the
sub-critical two phase regime of the decorated lattice gas model. An
examination is made of the effect of applying field mixing
transformations to the coexistence density and energy density
distributions.  For sub-critical temperatures down to approximately
$0.9T_c$, it is found that the observed asymmetries of the coexistence
density distributions are well accounted for by the linear mixing of
the energy density into the ordering operator.

Finally we apply the techniques developed in the context of the
decorated lattice gas model, to a more realistic system, namely a 3D
polymer model. Simulation studies of the bond fluctuation model within
the grand canonical ensemble are used to obtain the joint distribution
of density and energy on the liquid vapour coexistence curve.  The
critical temperature, chemical potential and field mixing parameters
of the model are accurately determined by requiring the collapse of
the measured scaling operator distributions onto their known universal
fixed point forms. In the sub-critical region close to the critical
point, the observed asymmetries of the density distribution are again
found to be well described by field mixing transformations.

\section{Background}
\label{sec:back}

In this section we provide a brief overview of the principal features
of the mixed field FSS theory of reference \cite{WILDING1}, placing it
within the context of the present work.

The systems we consider are assumed to be contained in a volume $L^d$
(with $d=3$ in the simulations described below) and thermodynamically
open so that the particle number can fluctuate. The observables on
which we shall focus are the particle number density:

\begin{equation}
\rho =L^{-d}N
\label{eq:phi}
\end{equation}
and the dimensionless energy density:
\begin{equation}
u=L^{-d}w^{-1}\Phi(\{{\bf r}\})
\label{eq:u}
\end{equation}
where $\Phi(\{{\bf r}\})$ is the configurational energy of the system
which we assume takes the general two-body form:

\begin{equation}
\Phi(\{{\bf r}\})=\sum_{i,j}\phi(|{\bf r_i}-{\bf r_j}|) ,
\end{equation}
with $\phi(r)$ the two-body interaction potential (e.g.
square-well or Lennard-Jones) whose associated coupling
strength (well-depth) we denote $w$.

Within the formalism of the grand canonical ensemble (GCE), the joint
distribution of density and energy fluctuations, $p_L(\rho,u)$, is
controlled by the reduced chemical potential $\mu$ and the coupling
strength $w$ (both in units of $k_BT$).  The critical point of the
system is located by critical values of the chemical potential $\mu_c$
and coupling $w_c$. Deviations of $w$ and $\mu$ from their critical
values control the sizes of the two relevant scaling field that
characterise the critical behaviour \cite{WEGNER}. In the absence of
the special symmetry prevailing in the Ising model, the relevant
scaling fields comprise (asymptotically) {\em linear combinations} of
the coupling and chemical potential difference \cite{REHR}:

\begin{equation}
\tau = w_c-w+s(\mu - \mu_c) \hspace{1cm} h=\mu - \mu_c+ r(w_c-w)
\label{eq:scaflds}
\end{equation}
where $\tau$ is the thermal scaling field and $h$ is the
ordering scaling field.  The parameters \mix\ and \mixp\ are
system-specific quantities controlling the degree of field mixing. In
particular $\mixp$ is identifiable as the limiting critical gradient
of the coexistence curve in
the space of $\mu$ and $w$. The role of $s$ is somewhat
less tangible; it controls the degree to which the chemical potential
features in the thermal scaling field, manifest in
the widely observed critical singularity of the coexistence curve
diameter of fluids \cite{SENGERS}.

Conjugate to the two relevant scaling fields are scaling operators
\oM\ and \oE , which comprise linear combinations of the particle
density and energy density \cite{BRUCE2,WILDING1}:

\begin{equation}
\oM  = \dfac \left[ \rho - s u \right] \hspace{1cm} \oE  =  \dfac \left[  u  -
r \rho \right]
\label{eq:oplinks}
\end{equation}
The operator \oM\ (which is conjugate to the ordering field $h$) is
termed the ordering operator, while \oE\ (conjugate to the
thermal field) is termed the energy-like operator.  In the
special case of models of the Ising symmetry, (for which $\mix = \mixp
=0$), \oM\ is simply the magnetisation while \oE\ is the energy
density.

The joint distribution of density and energy is simply related to the
joint distribution of mixed operators:

\begin{equation}
p_L(\rho,u) = \dfac p_L(\oM , \oE)
\label{eq:pdflink}
\end{equation}
Near criticality, and in the limit of large system size,
$p_L(\oM,\oE)$ is expected to be describable by a finite-size-scaling
relation of the form \cite{WILDING1}:

\setcounter{abc}{1}
\begin{equation}
\label{eq:ansatz}
p_L(\oM , \oE) \simeq  \LMp \LEp \tilde{p} (\LMp \delta \oM , \LEp \delta \oE ,
\LM \gM , \LE \gE )
\end{equation}
\addtocounter{equation}{-1}
\addtocounter{abc}{1}
where
\begin{equation}
\label{eq:Lamdefs}
\LE = \aE L^{1/\nu} \hspace{1cm} \LM = \aM L^{d-\beta/\nu}\hspace{1cm} \LM \LMp
= \LE \LEp = L^d
\end{equation}
\addtocounter{equation}{-1}
\addtocounter{abc}{1}
and
\begin{equation}
\label{eq:deltops}
\delta \oM \equiv  \oM - <\oM >_c  \hspace{1cm} \delta \oE \equiv  \oE - <\oE
>_c
\end{equation}
\setcounter{abc}{0}
The subscripts c in equations~\ref{eq:deltops} signify that the averages are to
be taken at criticality.  Given appropriate choices for the non-universal
scale factors \aM\ and \aE\ (equation~\ref{eq:Lamdefs}), the
function $\tilde{p}^\star$ is expected to be universal.

Precisely at criticality, equation~\ref{eq:ansatz} implies simply

\begin{equation}
p_L(\oM , \oE) \simeq  \LMp \LEp \tilde{p}^\star (\LMp \delta \oM , \LEp \delta
\oE)
\label{eq:critlim}
\end{equation}
where $\ptMEstar=\tilde{p}(x,y,0,0)$ is a function
describing the universal and statistically scale invariant operator
fluctuations characteristic of the critical point.

In what follows, we shall employ Monte Carlo simulations to explicitly
test the prediction of equation~\ref{eq:critlim} for a decorated
lattice gas model and a polymer system, both members of the Ising
universality class. To do so, however, we first require an independent
estimate of the fixed point function \ptMEstar\ appropriate to the
3D Ising universality class. In practice, this is most readily obtained
by considering the prototype member of the Ising class, namely the 3D
Ising model itself. Owing to its lack of field mixing, the scaling
operators of the Ising model are simply $\oM\rightarrow m$ (the
magnetisation) and $\oE\rightarrow u$ (the energy density). Moreover,
the availability of highly accurate estimates for the Ising model
critical temperature, circumvents the need to perform a time consuming
search for the critical point.

\section{Monte Carlo simulations}
\label{sec:mc}
\setcounter{equation}{0}
\setcounter{abc}{0}

\subsection{The 3D Ising model and the form of \ptMEstar\ }
\label{sec:ising}

Using a vectorised algorithm on a Cray YMP supercomputer, we have
performed high precision Monte Carlo simulation measurements of the
joint magnetisation and energy density distribution for the 3D Ising
model on a periodic lattice of side $L=20$. The measurements were
performed at the estimated (reduced) critical coupling
$K_c^I=0.2216595(26)$, as obtained in a previous high precision Monte
Carlo study \cite{FERRENBERG1}. Following an initial equilibration
period of $2\times10^6$ Monte Carlo steps per spin (MCS), the
magnetisation and energy density were sampled at intervals of $50$ MCS
(in order to reduce correlations), and the results accumulated in a
histogram. The final histogram (comprising $2\times10^7$ entries) was
formed from $12$ independent runs, thereby allowing the statistical
independence of the data to be assessed and statistical errors to be
assigned to the results.  The resulting form of \ptMEstar , normalised
to unit integrated weight and scaled to unit variance along both axes,
is shown in figure~\ref{fig:j_ising}. The associated ordering
operator distribution $\ptMstar=\int \ptMEstar dy$ , and energy
operator distribution $\ptEstar =\int \ptMEstar dx$ are shown in
figure~\ref{fig:i_m&u}. One observes that while the form of \ptMstar\
is doubly peaked and symmetric, that of \ptEstar\ is singly peaked and
asymmetric.

\subsection{The 3D decorated lattice gas model}
\label{sec:dlg_intro}

The decorated lattice gas model was first proposed in its two
dimensional form by Mermin as an example of a system exhibiting a
singular coexistence diameter \cite{MERMIN}. The model was
subsequently generalised to simple, body, and face centred cubic
lattices by other workers \cite{ZOLLWEG,MULHOLLAND1, MULHOLLAND2} and
studied for its interesting coexistence properties. The simple cubic
form of the model, on which we shall focus in the present work,
consists of an ordinary simple cubic lattice gas (whose sites we term
the primary sites), augmented (decorated) by additional secondary
sites on the intersitial bonds. Particles on primary sites are
assumed to interact with one another via a dimensionless coupling of
strength $\lambda$, while particles on the secondary sites interact
with those on primary sites via a dimensionless coupling $\eta$, but
do not interact with each other. A schematic representation of a unit
cell of the model is shown in figure~\ref{fig:dlatt}.

The configurational energy $\Phi(\{\sigma\})$ of the decorated lattice
gas model is given by

\begin{equation}
\Phi(\{\sigma\}) = \sum_{<i,j>}\eta\sigma_i\sigma_j +
\sum_{[m,n]}\lambda\sigma_m\sigma_n \label{eq:pot}
\label{eq:confen}
\end{equation}
with $\sigma_i=0,1$.  The site indices $i$ and $j$ are taken to run
over nearest neighbour primary and secondary sites of the model, while $m$ and
$n$ run only over nearest neighbour primary sites. It is straightforward to
show
that the particle-hole symmetry that obtains in the ordinary lattice
gas is equivalent to the requirement that all sites have the same
average energy environment.  The presence of two inequivalent
sublattices in the decorated model clearly violates this condition and
leads to field mixing.

Aside from its field mixing properties, the chief asset of the
decorated lattice gas model, is its analytic tractability.  The grand
partition function of the asymmetric model can be related by means of
analytic transforms to that of the ordinary lattice gas model which is
itself isomorphic to the Ising model. Specifically, one finds:

\begin{equation}
\Omega (\mu ,T)=(1+e^{\mu/kT})^{3{\overline N}}{\overline \Omega} ({\overline
\mu},{\overline T})
\label{eq:part}
\end{equation}
where $\Omega$ is the partition function of the decorated model and $\mu$
and $T$ are the chemical potential and temperature respectively.  Bars
denote quantities in the ordinary lattice gas and ${\overline N}$ is
the number of primary sites in the model.

Introducing the dimensionless chemical potential $\xi=\mu/k_bT$
equation~\ref{eq:part} leads to the following relationships
\cite{MULHOLLAND1}:

\setcounter{abc}{1}
\begin{eqnarray}
{\overline \xi} & = & \xi + 6\ln\left [ \frac{1+e^{\xi+
\eta}}{1+e^\xi}\right]\\
\addtocounter{equation}{-1}
\label{eq:relnsa}
\addtocounter{abc}{1}
{\overline \lambda} & = & \lambda + \ln \left [\frac{(1+e^\xi
)(1+e^{\xi+2\eta})}{(1+e^{\xi+\eta })^2}\right ]
\label{eq:relnsb}
\end{eqnarray}
\setcounter{abc}{0}
where ${\overline \xi}={\overline \mu}/k_b{\overline T}$ and ${\overline
\lambda}={\overline \lambda }/k_b{\overline T}$ are respectively the
dimensionless
chemical potential and dimensionless nearest neighbour coupling constant of
the ordinary lattice gas.

In the 3D ordinary lattice gas the liquid-vapour coexistence line is
specified by the condition ${\overline \xi}=-3{\overline \lambda}$
\cite{LEE}. The location of critical point that terminates this line
is not known exactly, but is trivially related to that of the Ising
model:

\begin{equation}
{\overline \lambda_c}=4K_c^I, \hspace{1cm} {\overline \xi}_c=-3{\overline
\lambda}_c
\label{eq:crittemp}
\end{equation}
where
\begin{equation}
K_c^I= 0.2216595(26)
\label{eq:crit_is}
\end{equation}
is the estimated dimensionless critical coupling of the simple cubic
Ising model \cite{FERRENBERG1}. Estimates for the critical point
parameters are thus obtainable by feeding this value of $K_c^I$ into
equations~\ref{eq:relnsa} and~\ref{eq:relnsb}. Similarly, the
coexistence curve of the decorated model is obtainable by setting
${\overline \xi}=-3{\overline \lambda}$ in equations~\ref{eq:relnsa}
and \ref{eq:relnsb} to find:

\begin{equation}
\xi + 3\lambda+3\ln\left [ \frac{1+e^{\xi + 2\eta}}{1+e^\xi}\right ] = 0
\label{eq:coexsur}
\end{equation}
Note however, that since $\lambda$ and $\eta$ both enter only as
multiplicative factors in the configurational energy
(equation~\ref{eq:confen}), the coexistence curve is uniquely
parameterised by the value of the coupling ratio $\lambda/\eta$.
Varying this ratio allows one to tune the degree of field mixing
\cite{WILDING2}.  Indeed for the special choice $\lambda/\eta=-1/3$,
the average energy environment is identical for atoms on both
sublattices of the model and particle-hole symmetry is restored.

Knowledge of the mapping between the decorated lattice gas and the
ordinary lattice gas also permits an analytic calculation of the field
mixing parameters \mixp\ and \mix\ for the model. The value of $r$ is
obtainable from equation~\ref{eq:coexsur} simply by calculating the
limiting critical gradient of the coexistence curve. A calculation of
\mix\ proceeds from the observation that in the ordinary lattice gas,
the field-like scaling field \gM\ coincides with the line ${\overline
\lambda}={\overline \lambda_c}$ in the space of ${\overline \xi }$ and
${\overline \lambda}$.  It follows that in the decorated lattice gas
model, the direction of \gM\ can be obtained from
equation~\ref{eq:relnsb} by setting ${\overline \lambda}={\overline
\lambda_c}$, $\eta=\eta(\lambda)$ and solving for $\lambda$.  The
value of \mix\ is then given by

\begin{equation}
\mix =\left ( \frac{\partial \lambda }{\partial \xi}\right )_c
\label{eq:s_calc}
\end{equation}
where the derivative is to be evaluated at criticality.

Compared to more realistic fluid models such as the Lennard-Jones
fluid, the great simplicity of the decorated lattice gas model renders
it highly computationally tractable.  Moreover the prior availability
of accurate values for the critical parameters obviates the need for a
time consuming search of parameter space for the critical point, and
simplifies the task of data analysis. The model therefore provides
an ideal test-bed for simulation studies of critical point field
mixing.

%********Point out that u is in units of eta.******

\subsubsection{The critical limit}
\label{sec:critlim}

Using a Metropolis algorithm within the grand canonical ensemble (GCE)
\cite{BINDER4}, we have performed detailed simulation measurements of
the joint density and energy distribution $p_L(\rho,u)$ of the
decorated lattice gas model, at the estimated critical parameters
obtained as described above.  All simulations were performed for a
choice of the coupling ratio $\lambda/\eta=0.1$, for which it is known
that the coexistence curve of the model closely resembles those of
many real fluids \cite{MULHOLLAND1}. In the course of the simulations,
three system sizes were studied having linear extent $L=12,L=20$ and
$L=32$. Periodic boundary conditions were employed throughout.  Prior
to data collection, equilibration periods of $5\times10^6$ MCS were
utilised.  Samples of the density and energy density were then
performed at intervals of $50$ MCS (to reduce correlations) and the
data stored in histograms. For each system size, $12$ independent runs
were performed in order to test the statistical independence of the
data and to assign statistical errors to the results.  The final
histograms of $p_L(\rho,u)$ comprised $2\times10^7$ entries for the
$L=12$ and $L=20$ system sizes, and $1\times10^7$ entries for the
$L=32$ system size.

The measured form of $p_L(\rho,u)$ for the decorated lattice gas model
is presented in figure~\ref{fig:j_latt} for the $L=20$ system size.
Clearly apart from a general overall double peaked structure, the form
of this distribution bears little resemblance to that of \ptMEstar\
(c.f. figure~\ref{fig:j_ising}) which represents the joint critical
order parameter and energy density distribution in the absence of
field mixing. To illustrate the differences it is instructive to
compare the fluid density distributions $p_L(\rho)=\int p_L(\rho,u)du$
with \ptMstar , and the fluid energy density distribution $p_L(u)=\int
p_L(\rho,u)d\rho$ with \ptEstar\ .

The forms of $p_L(\rho)$ for all three system sizes are shown in
figure~\ref{fig:l_bare}a. In contrast to \ptMstar\
(figure~\ref{fig:i_m&u}a), all the density distributions exhibit a
pronounced asymmetry, qualitatively similar in form to that observed
in the 2D version of the same model \cite{WILDING2}. Even more conspicuous,
however, are the differences between the finite-size forms of $p_L(u)$
(figure~\ref{fig:l_bare}b), and that of \ptEstar\
(figure~\ref{fig:i_m&u}b). Clearly while the latter is singly peaked,
the former are {\em doubly peaked}, with a form (were one to plot
$p_L(-u)$), reminiscent of the density distribution.  As we shall
show, the explanation of these differences is to be found in the field
mixing that manifests the lack of particle-hole symmetry in fluids.

In order to expose the universality linking the critical point of the
decorated lattice gas model to that of the Ising model, it is
necessary to recast $p_L(\rho,u)$ in terms of the joint distribution
of scaling operators \pLME\ , cf. equation~\ref{eq:critlim}. To do so
however, requires specification of the field mixing parameters \mix\
and \mixp\ featuring in the definitions of \oM\ and \oE\
(equation~\ref{eq:oplinks}). In practice, these values may be readily
found by requiring that the {\em single} scaling operators
distributions \pLM\ and \pLE\ match their respective fixed point forms
\ptMstar\ and \ptEstar\ .  Carrying out this procedure yields the
matchings shown in figure~\ref{fig:lat_ops}a and \ref{fig:lat_ops}b.
The associated estimates for the field mixing parameters are
$\mix=-0.143(8)$ and $\mixp=-3.11(1)$. These values compare very
favourably with those calculable analytically, for which one finds
$\mix=-0.1428\cdots , \mixp=-3.1163\cdots$ Such a high level of
accord indicates that the matching of the operator distributions to
the universal Ising forms is a potentially very accurate method for
determining the field mixing parameters.

Having obtained estimates for \mix\ and \mixp , one may then construct
the joint distribution of scaling operators \pLME . The resulting form
is shown in figure~\ref{fig:tjhist}, and should be compared with that
of \ptMEstar\ shown in figure~\ref{fig:j_ising}. Clearly the agreement
between the operator distributions and the universal fixed point form
is gratifying, providing substantial corroboration of the mixed field
FSS theory.

\subsubsection{The subcritical region}

As the critical point is approached along the line of phase
coexistence, the known symmetries of the Ising problem imply that
\setcounter{abc}{1}
\begin{eqnarray}
\langle \oM\rangle^\pm-\langle\oM\rangle_c &=& \pm a |\gE |^\beta ,\\
\addtocounter{equation}{-1}
\label{eq:approacha}
\addtocounter{abc}{1}
\label{eq:approachb}
\langle \oE\rangle^\pm-\langle\oE\rangle_c &=& b |\gE |^{1-\alpha} +
\mbox{terms analytic at criticality}
\end{eqnarray}
\setcounter{abc}{0}
where $a$ and $b$ are critical amplitudes and $\pm$ denote limits as
the coexistence curve is approached from above ($\gM\rightarrow 0^+$)
or below ($\gM\rightarrow0^-$). Recalling that $\oM=\rho+\mix\oE$ then
yields the two branches of the coexistence curve densities near
criticality:

\begin{equation}
\rho_\pm-\rho_c = \pm a|\gE|^\beta+sb |\gE |^{1-\alpha} +
\mbox{terms analytic at criticality} ,
\label{eq:coexcv}
\end{equation}
which displays a singular diameter:

\begin{equation}
\rho_d-\rho_c=\frac{1}{2}(\rho_++\rho_-)-\rho_c\sim sb |\gE |^{1-\alpha}
\end{equation}
as is indeed observed experimentally \cite{SENGERS}.

At temperatures outside the critical region, the
relations~\ref{eq:approacha} and ~\ref{eq:approachb} are {\em not}
generally expected to hold and a crossover description to regular
classical behaviour is more appropriate
\cite{SINGH,SENGERS2,SENGERS3,SENGERS4,SENGERS5,ZCHEN}.  Nevertheless
as observed in many computer simulations of various simple fluids, the
temperature dependence of the order parameter {\em is} quite
accurately described by Ising critical exponents over a remarkably
wide range of subcritical temperatures \cite{PANAGIO}, without the
need to introduce higher order (non-linear) terms in the Wegner
expansion of the scaling fields \cite{WEGNER}. In view of this it
seems of interest to examine the range of applicability of the scaling
form~\ref{eq:ansatz} in the subcritical two-phase region.

To this end we have obtained the joint density and energy density
distribution of the decorated lattice gas model at temperatures
$T=0.9T_c$ and $T=0.8T_c$, for a system of side $L=20$. In order to
circumvent the prohibitively large `tunneling' times between the
coexisting phases that normally plague GCE simulations in the
two-phase region, we have employed the multicanonical preweighting
scheme \cite{BERG}. This scheme uses weighted transitions to encourage
the system to sample those interfacial configurations that would
otherwise occur only very rarely.  The weight factors are
chosen such that the sampled density distribution is approximately
flat in the density region between the peaks of the two coexisting
phases.  After the simulation, the correct (`unweighted') coexistence
distribution is regained by dividing out the weight factors from
the sampled distribution. In this manner the effective tunneling
frequency between the coexisting phases may be increased by many
orders of magnitude, thereby facilitating very accurate estimates of
the coexistence form of $p_L(\rho,u)$.

Using the multicanonical preweighting scheme, samples of the density
and energy were accumulated every $20$ MCS and stored in
histograms. The final histograms for $p_L(\rho,u)$ comprised
approximately $5\times10^7$ entries. In figure~\ref{fig:coex_ops}a we
present the measured coexistence forms of $p_L(\rho)$ for $T=0.9T_c$
and $T=0.8T_c$. Also included in the figure is the critical point form
of $p_L(\rho)$ for the $L=20$ system size, as obtained
previously(c.f. subsection~\ref{sec:critlim}).  Clearly the
distributions are all asymmetric, those for the two subcritical
temperatures having a form similar to those observed in an asymmetric
version of the 2D Blume-Emery-Griffiths spin model
\cite{BORGS}. Although the two peaks of the sub-critical distributions
have equal weight, (reflecting the thermodynamic condition for
coexistence \cite{EWR}), one observes that the high density peak is
much broader and shorter than the low density peak. Moreover the
magnitude of the change in position of the high density peak on
lowering the temperature is much greater than that in the low density
peak.  It is this effect that gives rise to the asymmetric form of the
temperature-density phase diagram of fluids.

In figure~\ref{fig:coex_ops}b we present the corresponding forms of
the ordering operator distribution \pLM . For each temperature the
value of the field mixing parameter \mixp\ was obtained as the
gradient of the coexistence curve, while the value of \mix\ was
obtained by applying the prescription of equation~\ref{eq:s_calc}. One
sees that in this choice of variables, the distributions at $T=T_c$
and $T=0.9T_c$ are largely {\em symmetric} i.e. both peaks have the
same height and width. Only the distribution for $T=0.8T_c$ exhibits
small deviations from symmetry, these being attributable to the
non-singular (and non-Ising) part of the partition function
(equation~\ref{eq:part}).  Thus it would appear that in the present
case at least, the validity of the scaling form ~\ref{eq:ansatz}
extends some $10\%$ or more below the critical temperature. In this
temperature range, the asymmetry of the density distribution can
therefore be accurately ascribed to the linear mixing of the energy
density into the ordering operator. It remains to be seen however, to
what extent this finding holds true in more realistic fluid models as
well as those possessing very large coexistence curve asymmetries,
such as polar \cite{VANLEEUWEN}, ionic \cite{SENGERS1} or metallic
fluids \cite{JUNGST,GOLDSTEIN,GOLDSTEIN1}.

It is also instructive to compare the simulation results with analytic
calculations of the coexistence density diameter $\rho_d$ and the
ordering operator diameter $\oM_d$. The latter can be calculated
exactly, while the former may be obtained approximately by employing
the tabulated values of the Pad\'{e} Approximants for the temperature
dependence of the Ising model energy density \cite{SCESNEY}.
Performing these calculations for fractional temperatures $t=T/T_c$ in
the range $0.7\le t \le 1.0$ yields the results shown in
figure~\ref{fig:analyt}. The weak critical singularity in the
coexistence diameter is readily discernible from the figure although,
as expected, no such singularity obtains for $\oM_d(t)$, which is
analytic at the critical point. Some variation in $\oM_d$ is seen as a
function of temperature, however, but this again arises from the
non-singular, non-Ising part of the partition function~\ref{eq:part}.
Also included in the figure are the simulation estimates of
$\langle\oM\rangle (t)$ and $\langle\rho\rangle(t)$ obtained from the
multicanonical simulations at $T=0.9T_c$ and $T=0.8Tc$, as well as the
conventional simulations at the critical point.  Clearly a very good
overall agreement between the simulations and the analytical
predictions is apparent.

%Need to say we had to tune $\mu$ to get equal weights.

\subsection{The universal critical finite-size spectrum of \newline
$p_L(\rho)$ and $p_L(u)$}

In this subsection we return to a consideration of the critical point
forms of $p_L(\rho )$ and $p_L(u)$ in fluids, with the aim of gaining
an understanding of their shapes and finite-size-scaling behaviour. To
this end it is expedient to reexpress $\rho$ and $u$ in terms of the
scaling operators. Appealing to equation~\ref{eq:oplinks}, one finds

\begin{equation}
u=\oE-\mixp\oM \hspace{1cm} \rho=\oM - s \oE ,
\label{eq:udenmix}
\end{equation}
so that the critical density and energy density distributions are

\begin{equation}
p_L(u)=p_L (\oE -r\oM ) \hspace{1cm} p_L(\rho)=p_L (\oM -s\oE )
\label{eq:prho&u}
\end{equation}

Now the structure of the scaling form~\ref{eq:ansatz} shows that the
typical size of the fluctuations in the energy-like operator will vary
with system size like $\delta\oE\sim L^{-(1-\alpha)/\nu}$, while the
typical size of the fluctuations in the ordering operator vary like
$\delta\oM\sim L^{-\beta/\nu}$ . It follows that for a given $L$, the
shape of the energy and density distributions can be identified with
the distribution of the variable

\begin{equation}
X_{\Theta} = a_{\cal M}^{-1}\delta \oM \cos \Theta
+ a_{\cal E}^{-1}\delta\oE \sin \Theta  ,
\label{eq:Theta}
\end{equation}
with

\begin{equation}
\tan \Theta_{u} = \frac{a_{\oE}}{r a_{\oM}} L^{-(1-\alpha -\beta)/\nu}
\hspace{1cm} \mbox{and} \tan \Theta_{\rho} = \frac{s a_{\oE}}{a_{\oM}}
L^{-(1-\alpha -\beta)/\nu}
\label{eq:Ldep}
\end{equation}
where the subscripts $u$ and $\rho$ signify that the value of $\Theta$
corresponds to the energy density and density distributions
respectively.

The distributions $p(X_\Theta)$ constitute a one-parameter class of
{\em universal} functions describing the density and energy
distributions of fluids at finite $L$.  Geometrically, $\Theta$ can be
interpreted as defining a direction $OX_\Theta$ in the basal plane
formed by the $Ox$ and $Oy$ axes of figure~\ref{fig:j_ising}, making
an angle $\Theta$ with the $Ox$ axis. The form of $p(X_\Theta)$ is
then obtainable by projecting \ptMEstar , onto the vertical plane
which includes the line $OX_\Theta$. A representative selection of
such projections is shown in figure~\ref{fig:projs}. For
$\Theta=0^\circ$ one obtains simply the ordering operator distribution
\ptMstar\ , while for $\Theta=90^\circ$ the from is that of the energy-like
operator
\ptEstar\ distribution.  Intermediate between these values
a range of behaviour is obtained, representing the finite $L$ forms of
$p_L(\rho )$ and $p_L(u)$.

Asymptotically (i.e. as $L\rightarrow \infty$), equation~\ref{eq:Ldep}
implies that both $\Theta_u$ and $\Theta_\rho$ approach zero
\cite{FN2} so that in this limit

\begin{equation}
p_L(u)= p_L (-r\oM )\simeq\aM ^{-1}\mixp L^{\beta/\nu
}\tilde{p}_{\cal M}^\star (-\aM ^{-1}\mixp L^{\beta/\nu} \delta \oM )\\
\label{eq:lim_u}
\end{equation}

\begin{equation}
p_L(\rho)= p_L (\oM )\simeq\aM ^{-1} L^{\beta/\nu
}\tilde{p}_{\cal M}^\star (\aM ^{-1} L^{\beta/\nu} \delta \oM )\\
\label{eq:lim_rho}
\end{equation}
It follows that for {\em any} finite \mix\ and \mixp , the limiting
critical point forms of $p_L(\rho)$ and $p_L(u)$ both match the
critical ordering operator distribution \ptMstar . The approach to
this limiting behaviour is indeed clearly evident in the distributions
of figure~\ref{fig:l_bare}. We note however that the limiting form of
$p_L(u)$ differs radically from that of the Ising model where, owing
to the absence of field mixing, $\lim_{L\to\infty}p_L(u)=
\ptEstar$. The profound influence of field mixing on the critical
behaviour of fluids should therefore be apparent \cite{FN3}.

Finally in this subsection, we point out that precise knowledge of the
location of the critical point (obtained eg. from the data collapse of
the scaling operators onto their fixed point forms), does {\em not}
imply the possibility of directly measuring the infinite-volume
density and energy density. To appreciate this, recall that

\begin{equation}
\langle u \rangle_c=\langle\oE\rangle_c-\mixp\langle\oM\rangle_c
\hspace{1cm} \langle\rho\rangle_c=\langle\oM\rangle_c-\mix\langle\oE\rangle_c
\end{equation}
Now, while symmetry considerations dictate that the value of
$\langle\oM\rangle_c=\int \pLM d\oM$ is independent of system size, no
such symmetry condition pertains to \pLE , whose average value
$\langle\oE\rangle_c=\int d\oE\pLE$ at criticality is expected to vary
with system size like

\begin{equation}
\langle\oE\rangle_c(L)-\langle\oE\rangle_c(\infty ) \sim L^{-(1-\alpha)/\nu}
\end{equation}
It follows that in order to extract infinite volume estimates of
$\rho_c$ and $u_c$ from simulations at the critical point, it is
necessary to extrapolate data from a number of different system sizes
to the thermodynamic limit.  This procedure is illustrated in
figure~\ref{fig:Lshift} for the decorated lattice gas model, using
critical point data from the three system sizes $L=12,20,32$. The
measured values of $\langle \rho\rangle_c(L)$ and $\langle
u\rangle_c(L)$ are plotted against $L^{-(1-\alpha)/\nu}$. A least
squares fit to the data yields the infinite volume estimates
$\rho_c=0.3371(1)$ and $u_c=-0.8385(6)$. We remark that the existence
of these finite-size shifts imply that the equal weight criterion
\cite{EWR,BORGS} for the order parameter distribution, while correctly
identifying the coexistence curve in the subcritical regime, must fail close
to the critical point \cite{MUELLER}.

\subsection{Liquid-vapour equilibria of a polymer model}
\label{sec:poly}

We now apply the techniques developed in the foregoing sections to the
study of critical phenomena and phase coexistence in a more realistic
model fluid, namely a polymer system. The model we consider is the
bond fluctuation model, a coarse grained lattice-based polymer model
which combines the essential qualitative features of real polymer
systems--- monomer excluded volume and connectivity--- with
computational tractability. Within the framework of the model, each
monomer occupies a whole unit cell of a periodic simple cubic
lattice. Excluded volume interactions are catered for by requiring
that no lattice site can simultaneously be occupied by two monomer
corners.  Monomers along the polymer chains are connected by bond
vectors which can assume one of $108$ possible values, providing for
$87$ distinct bond angles and $5$ distinct bond lengths. For a more
detailed description of the model, the reader is referred to the
literature \cite{BFM}.

Using a grand canonical simulation algorithm, we have simulated chains
of length $N=20$ monomers, interacting via a short range square well
potential, the range of which was set at $\sqrt{6}$ (in units
of the lattice spacing). Chain insertions and deletions were
facilitated by use of the configurational bias Monte Carlo (CBMC)
method of Siepmann \cite{SIEPMANN}. The essential idea behind the CBMC
method is to improve the low acceptance rate associated with random
trial chain insertion, by `growing' chains of favourable energy into
the system. A bookkeeping scheme maintains a record of the statistical
bias associated with choosing favourable chain conformations, and this
bias is subsequently removed when the acceptance probability is
calculated. The CBMC technique has also recently been used in
conjunction with Gibbs ensemble Monte Carlo simulations of
liquid-vapour phase coexistence of off-lattice alkanes models
\cite{SIEPMANN1}.

The quantities measured in the simulations were the monomer density:

\begin{equation}
\rho=8nN/V
\end{equation}
and the dimensionless energy density:

\begin{equation}
u=8w^{-1}\Phi(\{r\})/V
\end{equation}
where n is the number of chains, $\Phi(\{r\})$ is the configurational
energy, $w$ is the well depth and $V$ is the system volume which was
set at $V=40^3$.  Here the factor of $8$ derives from the number of lattice
sites occupied by one monomer. In the course of the simulations,
measurements of $\rho$ and $u$ were performed at intervals of $5000$
chain insertion attempts and accumulated in the joint histogram
$p_L(\rho,u)$. The final histogram comprised some $3\times10^5$
entries.

In contrast to the decorated lattice gas model considered in previous
sections, the line of liquid-vapour phase coexistence is not known
{\em a-priori} for the polymer system and must therefore be identified
empirically. The precise location of the coexistence curve is
prescribed by the equal weight criterion for the two peaks of the
density distribution \cite{EWR}.  Unfortunately, the task of
identifying the coexistence curve using this criterion is an extremely
time consuming and computationally demanding one, since the density
distribution is generally very sensitive to small deviations from
coexistence. In practice, however, it suffices to obtain data for only
a few points close to the coexistence curve. The full coexistence
curve between these points can subsequently be constructed using
histogram reweighting techniques \cite{FERRENBERG,DEUTSCH3}. Provided
that the measured density distributions are doubly peaked and the
temperatures studied are not too widely separated, this technique
permits a very accurate determination of the coexistence curve locus.

Starting with an initial well-depth $w=0.569$, the approximate value
of the coexistence chemical potential was determined by tuning $\mu$
until the density distribution exhibited two peaks. Again the
multicanonical preweighting scheme \cite{BERG} was employed in order
to overcome the otherwise very large tunnelling times between the
coexisting phases. A histogram extrapolation based on this data was
then used to estimate the value of the coexistence chemical potential
for a well-depth $w=0.56$, which lies close to the critical
well-depth. A further long runs was carried out at this
near-coexistence point. By extrapolating the measured near-coexistence
histograms of $p_L(\rho,u)$ in conjunction with the equal weight
criterion, we were then able to construct a sizeable portion of the
coexistence curve (and its analytic extension). Representative forms
of the density distributions along the line of coexistence and its
analytic continuation are shown in figure~\ref{fig:poly_opcx}a. The
coexistence curve, expressed as a function of the well-depth $w$ and
chemical potential $\mu$ is shown in figure~\ref{fig:cxcurve}.

To locate the critical point along the line of phase coexistence, we
utilised the universal matching condition for the operator
distributions \pLM\ and \pLE\ . Again applying the histogram
reweighting technique, the well-depth, chemical potential and field
mixing parameters were tuned until the forms of \pLM\ and \pLE\ most
accurately matched the universal critical Ising forms of
figure~\ref{fig:i_m&u}. The results of performing this procedure are
shown in figure~\ref{fig:polyops}. Given that the system contains an
average of only about $100$ polymer chains, the quality of the data
collapse is remarkable. The mappings shown were effected for a choice
of the parameters

\begin{equation}
w_c=0.5584(1), \hspace{0.5cm}\mu_c=-5.16425(2),\hspace{0.5cm} s=-0.135(4),
\hspace{0.5cm}r=-2.55(2)
\end{equation}
where we have defined $\mu$ to be the chemical potential per monomer.  The
corresponding critical density and energy density distributions are
shown in figure~\ref{fig:polycrit}. They yield the (finite-size) estimates
$\rho_c=0.199(3)$ and $u_c=-0.304(4)$.

Turning lastly to the subcritical two-phase regime, we have again
considered the effect of forming linear combinations of the density
and energy density. Figure~\ref{fig:polyops}b shows the form of the
ordering operator distribution \pLM\ obtained at coexistence using the
histogram reweighting technique for the same values of $w$ shown in
figure~\ref{fig:polyops}a. In each instance the values of \mix\ was
chosen so that the two peaks of \pLM\ had both equal heights and
equal weights, while the value of \mixp\ was chosen so that \pLE\ was
singly peaked. As was the case for the decorated lattice gas model,
simple field mixing transformations also appear to account for the
sub-critical coexistence curve asymmetries of the polymer density
distribution, at least over the limited range of $w$ studied here.

\section{Conclusions}
\label{sec:disc}

In summary we have provided explicit demonstration of the field mixing
transformations that link the fluctuation spectra of the order
parameter and energy in the critical fluid to those of the critical
Ising magnet. The results serve to underline the profound influence of
field mixing on the non universal critical behaviour of fluids.  This
influence is manifest most notably as a finite-size shift to the
measured critical density, and as an alteration to the limiting (large
L) form of the critical energy distribution. Field mixing is also
found to account for the observed asymmetries of the coexistence
density distribution over a sizeable portion of the sub-critical
region.

With regard to the general computational issues raised in this study,
it has been seen that effecting the data collapse of the fluid scaling
operator distributions onto their (independently known) universal
fixed point forms, provides a very powerful method for accurately
locating the critical point and determining the field mixing
parameters of model fluids. This use of the scaling operator
distributions represents the natural extension to fluids of the order
parameter distribution method deployed so successfully in the study of
symmetric spin models.  Thus, in principle at least, there would
appear to be no barriers to attaining similar degrees of accuracy in
the study of critical fluids as has previously been achieved for
lattice spin systems.

The successes of the present work (and of an earlier FSS study of the
2D Lennard-Jones fluid \cite{WILDING1}) also attest to the utility of
the grand canonical ensemble for simulation studies of near-critical
fluids.  The benefits of this ensemble stem principally from the fact
that density fluctuations are observable on the scale of the system
size itself, thus freeing the method of the interfacial effects and
additional length scales that complicate use of the `sub-block'
finite-size scaling technique within the canonical (NVT) ensemble
\cite{ROVERE,ROVERE1}. High quality results can therefore be obtained
using comparatively much smaller system sizes, with concomitant
savings in computational effort.

The ability to perform a full finite size scaling analyses in the
near-critical region also represents an important advantage of the GCE
approach over the Gibbs ensemble Monte Carlo (GEMC) simulation
technique \cite{PANAGIO}. In the GEMC method, the fluctuating box size
seems to preclude a rigorous FSS analysis \cite{PANAGIO1,MON}, thus
seriously hindering the accurate location of the critical point. The
GEMC is, nevertheless, very efficient for locating the
temperature-density phase diagram in the sub-critical regime. For this
task, use of the bare GCE method is only feasible for temperatures
within a few percent of the critical temperature because the otherwise
high interfacial free energy results in prohibitively large
`tunneling' times between the coexisting phases. Nonetheless, as we
have shown, this problem is surmountable by combining the GCE with
recently developed multicanonical preweighting and histogram
reweighting techniques, thereby enabling accurate studies of the
coexistence density and energy fluctuations even well below the
critical temperature.

\subsection*{Acknowledgements}

The authors have benefitted from helpful discussions with B.A.
Berg, K. Binder and A.M. Ferrenberg.  NBW acknowledges the financial
support of a Max Planck fellowship from the Max Planck Institut
f\"{u}r Polymerforschung, Mainz. Part of the simulations described
here were performed on the CRAY-YMP computers at the HLRZ J\"{u}lich
and the RHRK Universit\"{a}t Kaiserslautern. Partial support from the
Deutsche Forschungsgemeinschaft (DFG) under grant number Bi314/3-2 is
also gratefully acknowledged.

\newpage

\begin{figure}[h]
\vspace*{0.5 in}
%\setlength{\epsfxsize}{20cm}
%\centerline{\mbox{\epsffile{./Figs/j_ising.eps}}}

\caption{Estimates of the fixed point form of the joint scaling
operator distribution \ptMEstar\ appropriate to the 3D Ising
universality class, obtained as the joint magnetisation and energy
density distribution of the $L=20$ periodic 3D Ising model at the
estimated critical coupling $K=0.2216595$. The data is expressed in
terms of the scaling variables $x=a_{\cal
M}^{-1}L^{\beta/\nu}(\oM-\oM_c)$ and $y=a_{\cal
E}^{-1}L^{(1-\alpha)/\nu}(\oE-\oE_c)$, with the scale factors $a_{\cal
E}^{-1}$ and $a_{\cal M}^{-1}$ chosen so that the distribution has
unit variance along both axes.}

\label{fig:j_ising}
\end{figure}

\begin{figure}[h]
\vspace*{-0.25 in}
%\setlength{\epsfxsize}{11.5cm}
%\centerline{\mbox{\epsffile{./Figs/p_im.eps}}}
%\centerline{\mbox{\epsffile{./Figs/p_iu.eps}}}

\caption{{\bf(a)} The ordering operator distribution \ptMstar\
appropriate to the 3D Ising universality class, obtained as the
magnetisation distribution of the $L=20$ 3D periodic Ising model at
the assigned value of the critical coupling $K_c^I=0.2216595$, and
expressed in terms of the scaling variable $x=a_{\cal
M}^{-1}L^{\beta/\nu}(\oM-\oM_c)$.  {\bf (b)} The corresponding energy
operator distribution \ptEstar , expressed in terms of the scaling
variable $y=a_{\cal E}^{-1}L^{(1-\alpha)/\nu}(\oE-\oE_c)$. In both cases,
the value of the non-universal scale factors $a_{\cal E}^{-1}$ and
$a_{\cal M}^{-1}$ were chosen so that the distributions have unit
variance. Statistical errors do not exceed the symbol sizes.}

\label{fig:i_m&u}
\end{figure}

\begin{figure}[h]
%\setlength{\epsfxsize}{15cm}
%\centerline{\mbox{\epsffile{./Figs/lattgas.eps}}}
\vspace*{0.5 in}
\caption{Schematic representation of a unit cell of the decorated lattice
gas model. The primary sites have been shaded, while the secondary
(decoration) site have not.  Primary sites interact with other nearest
neighbour primary sites with a coupling $\lambda$, and with nearest
neighbour secondary sites with a coupling $\eta$.}

\label{fig:dlatt}
\end{figure}

\begin{figure}[h]
\vspace*{0.5 in}
%\setlength{\epsfxsize}{20cm}
%\centerline{\mbox{\epsffile{./Figs/j_lattgas.eps}}}

\caption{Estimates of the normalised joint distribution of density and
energy density $p_L(\rho,u)$ of the 3D decorated lattice gas model for
$L=20$, at the assigned value of the critical parameters. }

\label{fig:j_latt}
\end{figure}

\begin{figure}[h]
%\vspace*{0.5 in}
%\setlength{\epsfxsize}{12cm}
%\centerline{\mbox{\epsffile{./Figs/latt_nhist.eps}}}
%\centerline{\mbox{\epsffile{./Figs/latt_enhist.eps}}}

\caption{{\bf (a)} The density distributions of the 3D decorated
lattice gas model at the assigned values of the critical parameters,
for system sizes $L=12,L=20$ and $L=32$. {\bf (b)} The corresponding
energy density distributions. Statistical errors do not exceed the
symbol sizes.}

\label{fig:l_bare}
\end{figure}

\begin{figure}[h]
%\vspace*{0.5 in}
%\setlength{\epsfxsize}{11.5cm}
%\centerline{\mbox{\epsffile{./Figs/latt_oM.eps}}}
%\centerline{\mbox{\epsffile{./Figs/latt_oE.eps}}}

\caption{{\bf (a)} The critical ordering operator distribution \pLM\
of the decorated lattice gas model for the system sizes $L=20$ and
$L=32$. The $L=12$ data has been omitted for clarity. Also shown for
comparison is the universal ordering operator distribution \ptMstar\
(c.f.  figure~\protect\ref{fig:i_m&u}a). {\bf (b)} The critical energy
operator distribution \pLE\ compared with the universal fixed point
form \ptEstar\ (figure~\protect\ref{fig:i_m&u}b).  In both cases, the
non-universal scale factors $a_{\cal E}^{-1}$ and $a_{\cal M}^{-1}$
have been chosen to give unit variance for the $L=32$ data set. The
exponent ratios were taken to be $\beta/\nu=0.5176$ and
$\alpha/\nu=0.177$. Statistical errors do not exceed the system
sizes.}

\label{fig:lat_ops}
\end{figure}

\begin{figure}[h]
%\vspace*{0.5 in}
%\setlength{\epsfxsize}{20cm}
%\centerline{\mbox{\epsffile{./Figs/j_lattop.eps}}}
%\include{Figs/j_lattop}

\caption{The joint critical distribution of scaling operators \pLME\
for the 3D decorated lattice gas model, obtained by applying
appropriate field mixing transformation to the distribution of
figure~\protect\ref{fig:j_latt}. The data is expressed in terms of the
scaling variables $x=a_{\cal M}^{-1}L^{\beta/\nu}(\oM-\oM_c)$ and
$y=a_{\cal E}^{-1}L^{(1-\alpha)/\nu}(\oE-\oE_c)$. The values of the
non-universal scale factors $a_{\cal E}^{-1}$ and $a_{\cal M}^{-1}$
have been chosen to give unit variance along both axes.}

\label{fig:tjhist}
\end{figure}

\begin{figure}[h]
\vspace*{0.25 in}
%\setlength{\epsfxsize}{12cm}
%\centerline{\mbox{\epsffile{./Figs/nhist_cx.eps}}}
%\centerline{\mbox{\epsffile{./Figs/poM_cx.eps}}}

\caption{{\bf (a)} Estimates of the coexistence density distributions of
the $L=20$ decorated lattice gas model at $T=T_c$, $T=0.9T_c$ and
$T=0.8T_c$. {\bf (b)} The corresponding forms of the ordering operator
distribution $p_L(\oM)$. Statistical errors do not exceed the symbol sizes. }

\label{fig:coex_ops}
\end{figure}

\begin{figure}[h]
\vspace*{0.5 in}
%\setlength{\epsfxsize}{15cm}
%\centerline{\mbox{\epsffile{./Figs/analyt.eps}}}

\caption{Comparison of the calculated (lines) and measured (points)
temperature dependence of the coexistence density and ordering
operator diameters. The calculated temperature dependence was obtained
from exact and series expansion methods, as described in the text. The
data points were obtained from the simulations, except the estimate
for $\rho_c$, which derives from the finite-size extrapolation of
figure~\protect\ref{fig:Lshift}.  Statistical errors do not exceed the
symbol sizes.}

\label{fig:analyt}
\end{figure}

\begin{figure}[h]
\vspace*{0.5 in}
%\setlength{\epsfxsize}{15cm}
%\centerline{\mbox{\epsffile{./Figs/projs.eps}}}

\caption{Selections from the universal finite-size spectrum of
critical density and energy density distributions of fluids. The
distributions were obtained according to the procedure described in
the text.  The values of the non-universal scale factors $a_{\cal
E}^{-1}$ and $a_{\cal M}^{-1}$ have been chosen to ensure that the
distributions have unit variance}

\label{fig:projs}
\end{figure}

\begin{figure}[h]
%\vspace*{0.5 in}
%\setlength{\epsfxsize}{12cm}
%\centerline{\mbox{\epsffile{./Figs/avrho.eps}}}
%\centerline{\mbox{\epsffile{./Figs/aven.eps}}}

\caption{{\bf (a)} The measured average density
$\langle\rho\rangle_c(L)$ of the decorated lattice gas model at the
critical point, expressed as a function of $L^{-(1-\alpha)/\nu}$ . The
least-squares fit yields an infinite volume estimate
$\rho_c=0.3371(1)$. {\bf (b)} The measured average energy density
$\langle u\rangle_c(L)$ of the decorated lattice gas model at the
critical point, expressed as a function of $L^{-(1-\alpha)/\nu}$ . The
least-squares fit yields an infinite volume estimate
$u_c=-0.8385(6)$. }

\label{fig:Lshift}
\end{figure}

\begin{figure}[h]
\vspace*{0.25 in}
%\setlength{\epsfxsize}{12cm}
%\centerline{\mbox{\epsffile{./Figs/poly_ncx.eps}}}
%\centerline{\mbox{\epsffile{./Figs/poly_opcx.eps}}}

\caption{{\bf (a)} The monomer density distribution for a selection of
well-depths $w$ along the line of liquid-vapour
coexistence. The distributions were obtained by applying the histogram
reweighting technique to measured near-coexistence data for $w=0.56$
and $w=0.569$. {\bf (b)} The corresponding distributions of the
ordering operator \pLM\ .}

\label{fig:poly_opcx}
\end{figure}

\begin{figure}[h]
\vspace*{0.5 in}
%\setlength{\epsfxsize}{13cm}
%\centerline{\mbox{\epsffile{./Figs/cxcurve.eps}}}

\caption{The line of liquid-vapour phase coexistence and its analytic
extension in the space of $\mu$ and $w$, for values of $w$ in the range
$0.555-0.575$. The results were obtained by implementing the equal
weight criterion for the density distribution. Also shown are the
measured directions of the relevant scaling fields.}

\label{fig:cxcurve}
\end{figure}

\begin{figure}[h]
%\vspace*{0.5 in}
%\setlength{\epsfxsize}{12cm}
%\centerline{\mbox{\epsffile{./Figs/poM_poly.eps}}}
%\centerline{\mbox{\epsffile{./Figs/poE_poly.eps}}}

\caption{{\bf (a)} The ordering operator distribution \pLM\ of the
polymer model at the assigned critical parameters. Also shown for
comparison is the universal fixed point form \ptMstar\
(cf. figure~\protect\ref{fig:i_m&u}a).  {\bf (b)} The energy-like
operator distribution \pLE\ compared with the universal fixed point
form \ptEstar\ (cf. figure~\protect\ref{fig:i_m&u}b).  In both cases the
non-universal scale factors $a_{\cal E}^{-1}$ and $a_{\cal M}^{-1}$
have been chosen to give unit variance. Statistical errors are
indicated by representative error bars.}

\label{fig:polyops}
\end{figure}

\begin{figure}[h]
%\vspace*{0.5 in}
%\setlength{\epsfxsize}{12.5cm}
%\centerline{\mbox{\epsffile{./Figs/nhist_poly.eps}}}
%\centerline{\mbox{\epsffile{./Figs/enhist_poly.eps}}}

\caption{{\bf (a)} Estimates for the polymer monomer density distribution
$p_L(\rho)$ at the assigned values of the critical well-depth and
chemical potential.  {\bf (b)} Corresponding estimates for the polymer
energy density distribution $p_L(u)$.  Representative error bars are shown.}

\label{fig:polycrit}
\end{figure}


\begin{thebibliography}{50}
\bibitem{FN1} Following convention we employ the language
of the ordinary lattice gas model, which is formally equivalent to the
Ising model.
\bibitem{FN3} The differences between the energy distributions of
fluids and symmetric systems have also recently been discussed by us
elsewhere in the context of the critical finite-size scaling behaviour
of the heat capacity in a binary polymer mixture
\protect\cite{MUELLER}.
\end{thebibliography}
\end{document}